\documentclass[12pt]{article}
\usepackage[margin=1in]{geometry}
\usepackage[compact]{titlesec}  
\usepackage{multirow}

\usepackage{amsmath}
\usepackage{graphicx}
\usepackage{pdfpages}
\usepackage{booktabs}
\usepackage{hyperref}

\begin{document}

\title{A spatial variance-smoothing area level model for small area estimation of demographic rates}
\author{Peter A. Gao and Jon Wakefield}
\maketitle
\begin{abstract}
	
	Accurate estimates of subnational health and demographic indicators are critical for informing health policy decisions.  Many countries collect relevant data using complex household surveys, but when data are limited, direct survey weighted estimates of small area proportions may be unreliable. Area level models treating these direct estimates as response data can improve precision but often require known sampling variances of the direct estimators for all areas. In practice, the sampling variances are typically estimated, so standard approaches do not account for a key source of uncertainty. In order to account for variability in the estimated sampling variances, we propose a hierarchical Bayesian spatial area level model that smooths both the estimated means and sampling variances to produce point and interval estimates of small area proportions. Our model explicitly targets estimation of small area proportions rather than means of continuous variables and we consider examples of both moderate and low prevalence events. We demonstrate the performance of our approach via simulation and application to vaccination coverage and HIV prevalence data from the Demographic and Health Surveys.
	
\end{abstract}
\maketitle
\section{Introduction}
Subnational estimates of health and demographic indicators such as vaccination rates or neonatal mortality rates are critical for highlighting disparities and guiding policy interventions. Data on key health outcomes are often collected using national household surveys designed to produce reliable direct weighted estimators for national and regional demographic rates of interest. However, often estimates are desired for smaller subregions for which direct weighted estimation may be insufficiently precise. When data are limited, model-based small area estimation methods can often improve estimates by sharing information between areas. Among model-based methods, unit level approaches are popular in demographic research as they explicitly model individual survey responses and can incorporate available individual level covariate information. However, if care is not taken to account for informative sampling or other features of the survey design, estimates from unit level models may be biased or improperly calibrated \cite{parker_computationally_2022, fuglstad_two_2022}. On the other hand, area level model-based approaches model direct survey weighted estimates as noisy response data, using random effects to produce smoothed small area estimates. The Fay-Herriot area level model \cite{fay_estimates_1979}  assumes that for each area, the direct estimator $\widehat{p}_a$ is available and can be modeled using the Gaussian distribution centered around the finite population mean $p_a$:
\begin{equation}\label{eqn:msamp}
\widehat{p}_a\mid p_a, V_a\sim N(p_a, V_a)
\end{equation}
where $V_a$ denotes sampling variances, which are typically assumed to be known. This sampling model is combined with a so-called linking model for the finite population means:
\begin{equation}
p_a\mid \boldsymbol\beta, \sigma_u^2\sim N(\mathbf{x}_a^T\boldsymbol\beta, \sigma_u^2)
\end{equation}
where $\mathbf{x}_a$ denotes a vector of area level covariate values for area $a$ and $\boldsymbol\beta$ denotes the corresponding coefficients. The linking model variance $\sigma_u^2$ controls the magnitude of deviations from the mean model. Since the direct estimators $\widehat{p}_a$ account for the survey design via the use of survey weights, area level methods are less sensitive to the effects of informative sampling and other design features than unit level methods. Under certain regularity conditions, the resulting estimators are design consistent; for a review, see Rao and Molina (2015) \cite{rao_small_2015}.

In practice, the variances $V_a$ are usually estimated using sample-based estimators $\widehat{V}_a$, but the standard Fay-Herriot model does not account for uncertainty in $\widehat{V}_a$. This is a well known problem in the small area estimation literature \cite{arora_superiority_1997, rivest_mean_2002, bell_examining_2008} and has motivated a number of proposed extensions of Fay-Herriot that incorporate variance modeling. However, existing approaches often rely on the availability of informative area level covariates for modeling the variance of survey estimators and do not account for uncertainty in the modeled variance estimates, simply treating them as known and using them to replace the direct variance estimates in the standard Fay-Herriot model. To account for this source of  uncertainty, we propose a fully Bayesian area level model for small area proportions that jointly models the direct estimators and associated variance estimators. Our approach assumes that area level random effects are spatially correlated and induces spatial smoothing of both means and variances. While many existing extensions to Fay-Herriot focus on estimation of means of continuous variables, our method is designed for estimation of small area proportions. In health and demographic applications, such proportions are the most common targets of inference. In simulations, we find that our proposed method produces interval estimates that may help correct for undercoverage observed for the standard Fay-Herriot approach.

Below we expand upon our proposed model and draw connections to existing approaches. In Section \ref{s:motive}, we describe two examples of estimation problems involving estimation of subnational demographic rates to provide motivation for our new approach. Section \ref{s:review} reviews existing area level models and discusses recent efforts to incorporate variance smoothing for small area estimation. In Section \ref{s:methods}, we outline our new spatial variance-smoothing area level model for estimation of small area proportions.  We compare our approach with other area level methods via simulation in Section \ref{s:sim} and by application to data from the Demographic and Health Surveys in Section \ref{s:app}. Finally, we compare our method with existing approaches and identify potential directions for future research in Section \ref{s:dis}.

\section{Motivation}\label{s:motive}
\begin{figure}
	\centering
	\includegraphics[scale=.5]{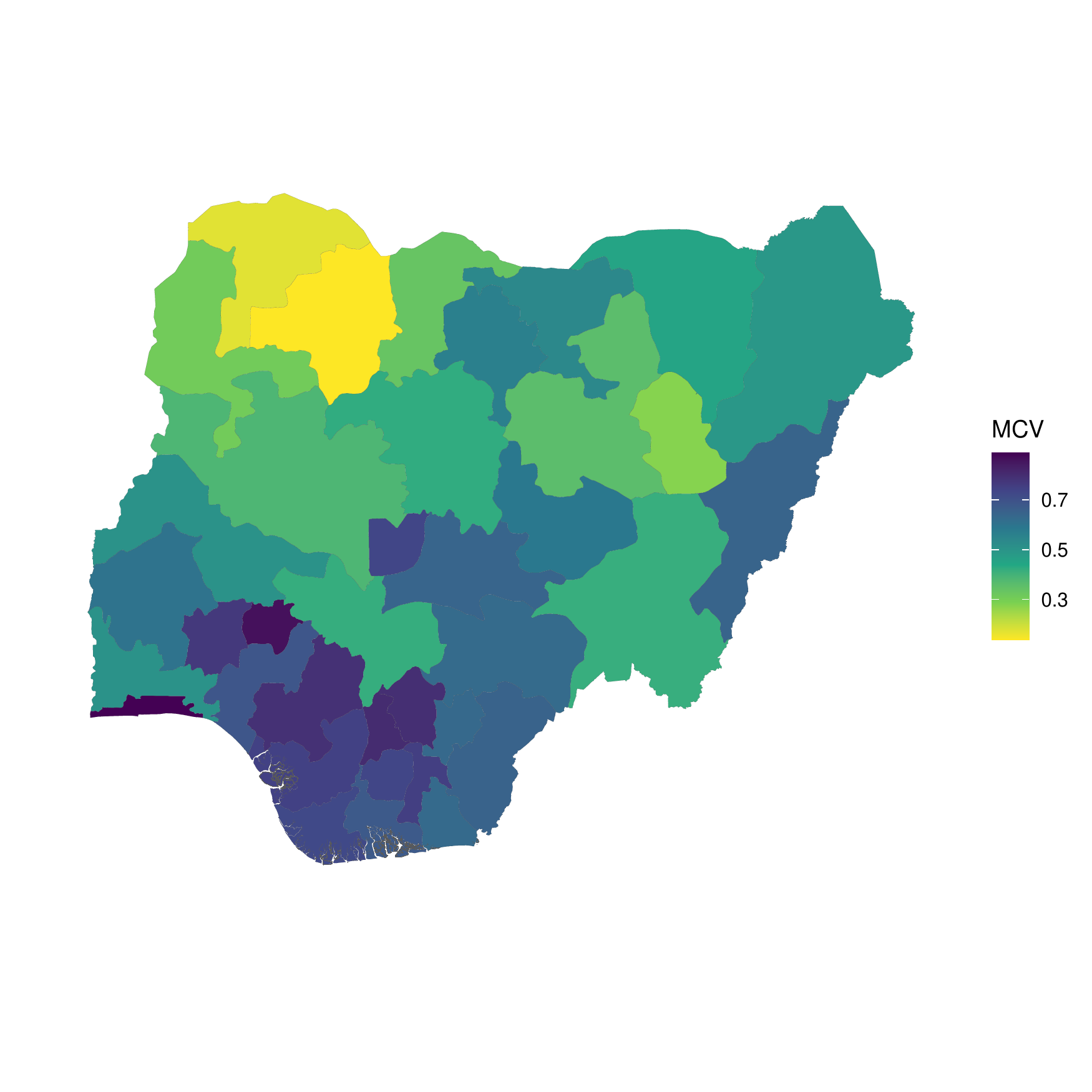}
	\includegraphics[scale=.5]{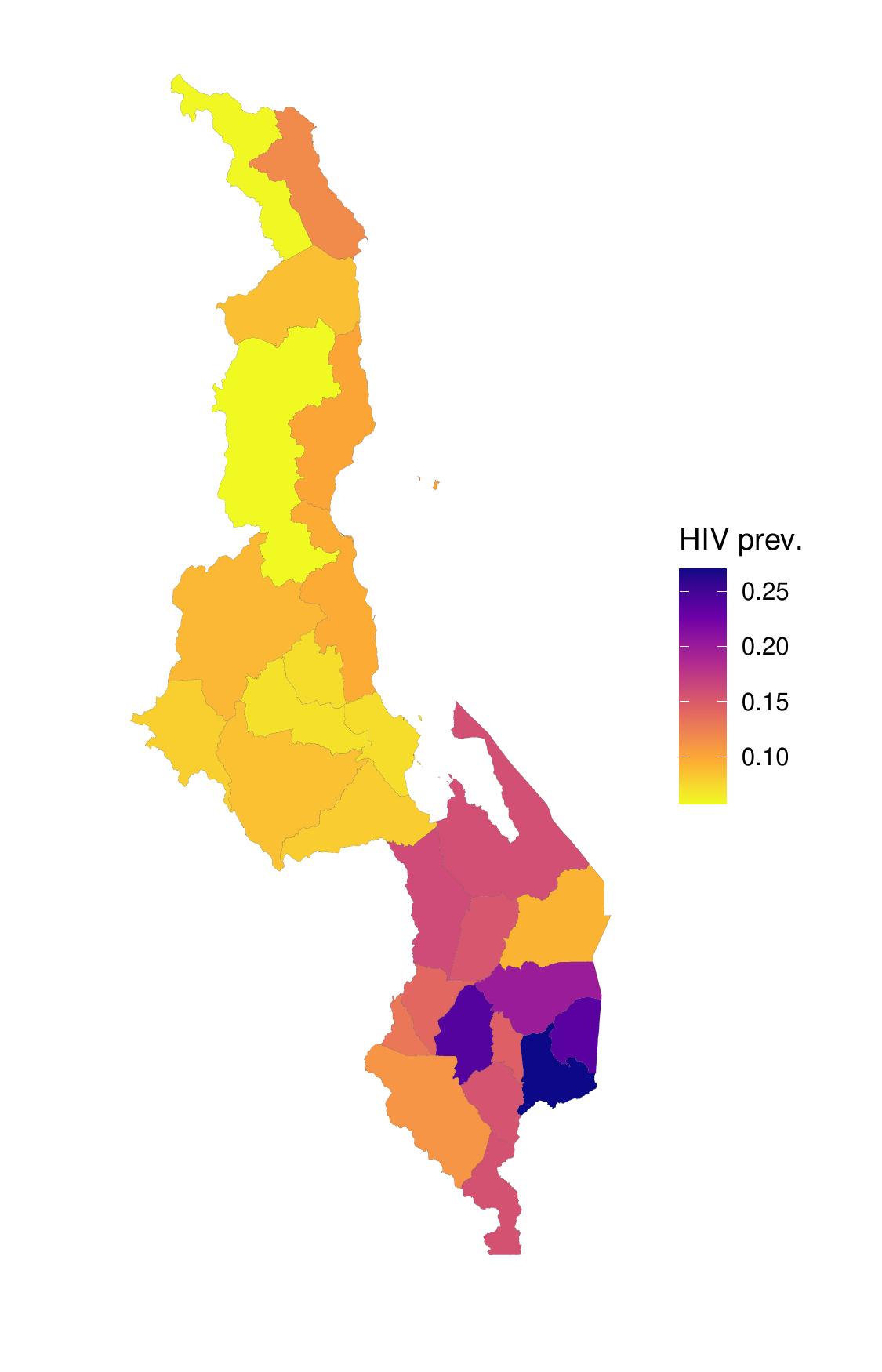}
	\caption{Direct weighted estimates of vaccination coverage rate for first dose of measles-containing-vaccine (MCV1) among children aged 12–23 months in Nigeria, 2018 (left) and HIV prevalence rate for women aged 15-49 in Malawi, 2015-2016 (right)}
	\label{fig:map-hajek}
\end{figure}

We consider two motivating examples of estimating subnational demographic rates using data from the Demographic and Health Surveys (DHS) Program. In the first, we use 2018 Nigeria DHS data to estimate regional vaccination coverage rates for the first dose of measles-containing-vaccine (MCV1) among children aged 12–23 months \cite{national_population_commission_-_npcnigeria_and_icf_nigeria_2019}. In the second, we use 2015-16 Malawi DHS data to estimate HIV prevalence among women aged 15–49 \cite{national_statistical_office_-_nsomalawi_and_icf_malawi_2017}. Figure \ref{fig:map-hajek} provides maps of direct survey-weighted estimators for both indicators. The measles vaccination example represents an estimation problem where the estimated area level proportions have a large spread and are generally located away from zero or one; in the HIV prevalence example, the direct estimates exhibit less variability and are on average closer to zero.

In Nigeria and Malawi, the DHS Program uses a stratified two-stage cluster sampling design. Countries are divided into administrative regions which are further partitioned into urban and rural areas. The sampling strata are defined by crossing these regions with urban/rural status. In Nigeria, the divisions used for defining strata are called Admin-1 regions; in Malawi, they are called Admin-2 regions. Each stratum is divided into collections of households called enumeration areas (EAs) or clusters. The first stage of sampling selects a pre-specified number of EAs in each stratum with probability proportional to the number of households in the EA. The second stage of sampling selects a fixed number of households in each sampled EA.

The 2018 Nigeria DHS collected data on measles vaccination status for children in sampled households based on vaccination cards or caregiver recall. In Nigeria, the Admin-1 regions consist of 36 states and the Federal Capital Territory of Abuja. For our analysis, we adhere to the Database of Global Administrative Areas (GADM) boundaries \\(https://gadm.org/download\_country\_v3.html). The sampling frame used for the Nigeria DHS was based on a 2006 national census which identified 664,999 EAs. Data were successfully collected in 1389 EAs, but due to security issues during the survey, a number of EAs were dropped. In particular, estimates for the Admin-1 area of Borno may have been affected (see Appendix A.3 of \cite{national_population_commission_-_npcnigeria_and_icf_nigeria_2019}). 

The 2015-16 Malawi DHS used  voluntary finger prick blood sampling to collect data on HIV prevalence. We desire estimates of HIV prevalence for each of Malawi's 28 districts, also referred to as Admin-2 areas. For this survey, the sampling frame was obtained from a 2008 census which identified 12,558 EAs distributed between 56 strata. Ultimately, data were collected from 827 EAs, from which a total of 8,497 women aged 15-49 were eligible for HIV testing. Ultimately, 93\% of eligible women were tested, but the HIV test results were anonymized, with volunteers not informed of their results and instead receiving access to educational materials and free counseling and testing \cite{national_statistical_office_-_nsomalawi_and_icf_malawi_2017}.

\begin{figure}
	\centering
	\includegraphics[scale=.5]{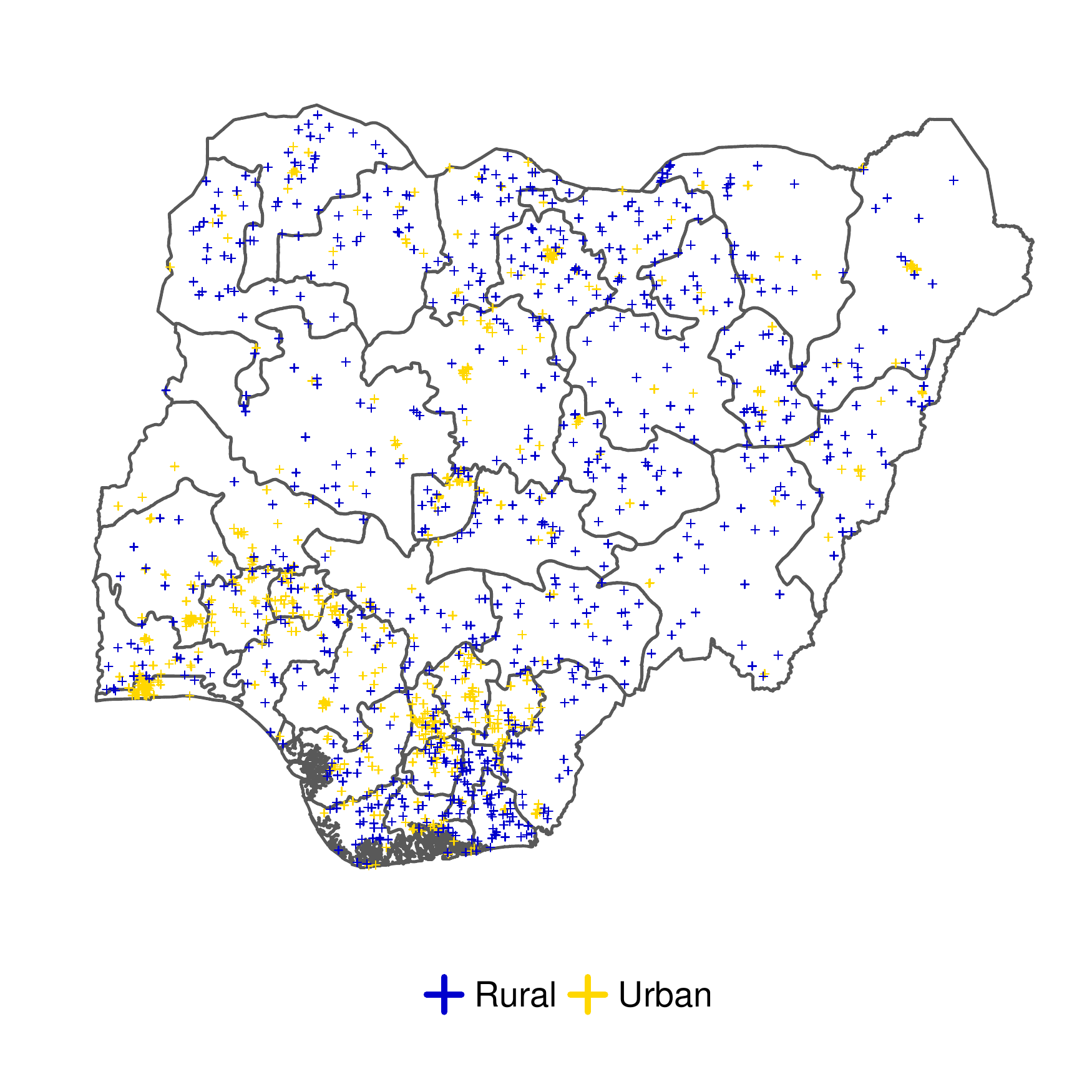}
	\includegraphics[scale=.5]{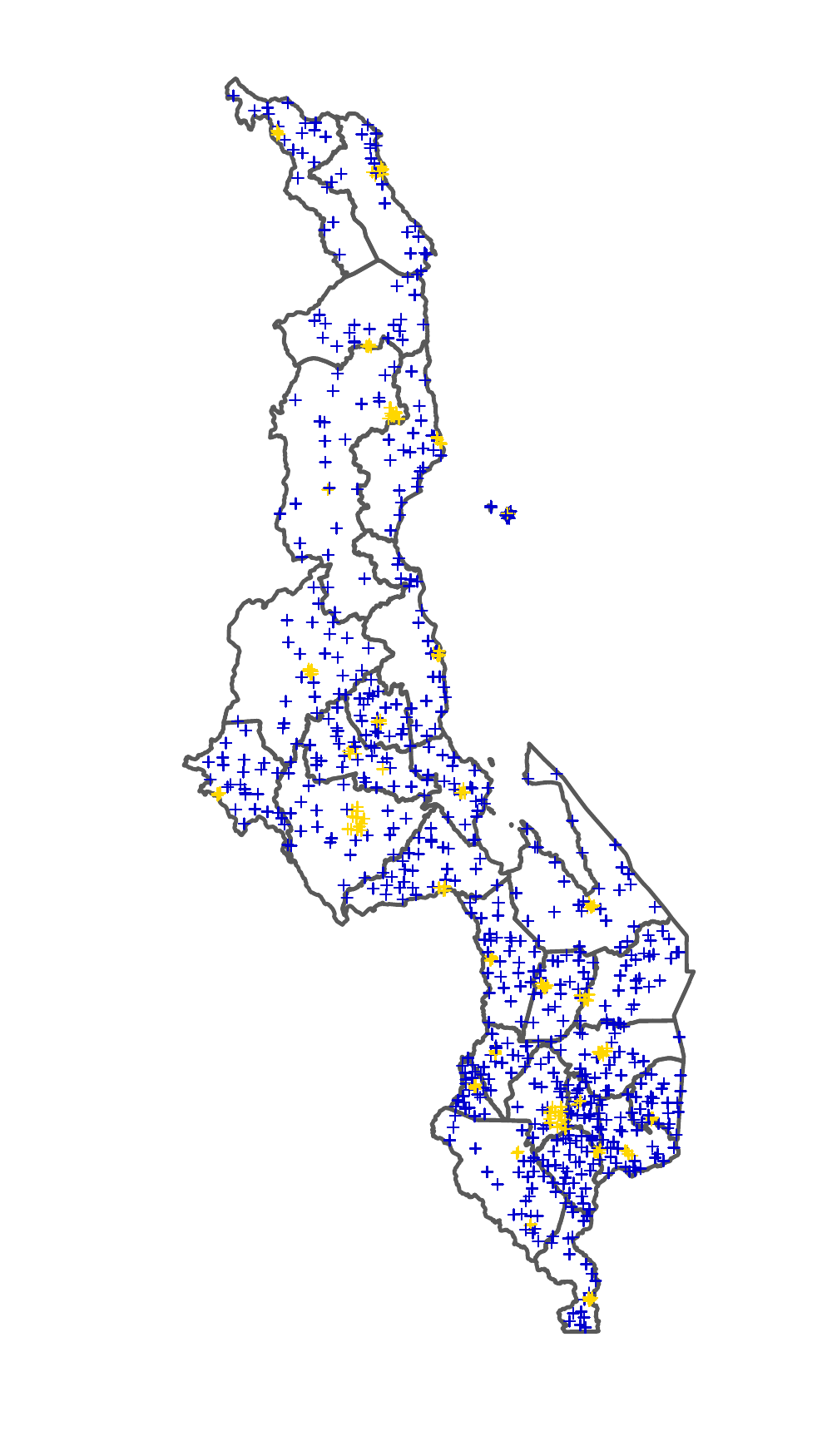}
	\caption{Small area boundaries and sampled enumeration area locations for Nigeria (left) and Malawi (right)}
	\label{fig:map}
\end{figure}

For both Nigeria and Malawi, the DHS provides GPS coordinates for nearly all EAs, but the locations have been adjusted to maintain privacy by adding small distances at random. Figure \ref{fig:map} provides maps of the small area boundaries and sampled EA locations in Nigeria and Malawi. In Malawi, since the island region of Likoma is disconnected from the mainland, we omit its data from our analysis.

\section{Existing Work}\label{s:review}

\subsection{Variance Smoothing for Continuous Response}

Standard area level models, including the Fay-Herriot model, do not account for uncertainty in the estimates of sampling variance $\widehat{V}_a$. As such, the basic Fay-Herriot model for continuous response data has been extended to account for unknown $V_a$ in a number of ways as introduced by Kleffe and Rao (1992) and Arora and Lahiri (1997)\cite{kleffe_estimation_1992, arora_superiority_1997} and  reviewed by Bell (2008) \cite{bell_examining_2008}. Below, we review extensions of area level models. Following Rivest and Vandal (2002) \cite{rivest_mean_2002} and Wang and Fuller (2003) \cite{wang_mean_2003}, You and Chapman (2006) \cite{you_small_2006} assume the following sampling model for the variance estimators $\widehat{V}_a$:
\begin{equation}\label{eqn:vsamp}
\widehat{V}_a\mid V_a\sim \frac{V_a}{d_a}\chi^2_{d_a}
\end{equation}
where $d_a$ denotes the degrees of freedom for area $a$. In addition, $\widehat{V}_a$ are assumed to be independent of the mean estimators $p_a$. If the response values for area $a$ were independently and identically distributed Gaussian random variables, the above model (\ref{eqn:vsamp}) would hold for the variance estimator $\widehat{V}_a=s_a^2/n_a$ with $d_a=n_a-1$,  where $n_a$ denotes the sample size for area $a$. When responses are sampled at random with replacement within areas, such an assumption may be appropriate, but for complicated sampling schemes, different values of $d_a$ or even alternative models may be necessary. You and Chapman (2006) \cite{you_small_2006} adopted a hierarchical Bayesian approach and placed inverse Gamma priors on the variance parameters $\sigma_u^2\sim IG(r_0, s_0)$ and $V_a\sim IG(r_a, s_a)$, with $r_a, s_a$ chosen to be small for all areas $a=1,\ldots A$. Notably, they allow the prior for $V_a$ to vary across areas, which makes the $V_a$ values independent across areas. Maiti et al. (2014) \cite{maiti_prediction_2014} assume the same variance sampling model (\ref{eqn:vsamp}), but adopt an empirical Bayes approach, setting the prior $\sigma_a^2\sim IG(r, s)$ and estimating $\{r, s, \sigma_u^2\}$ via maximum likelihood.

In addition to modeling $\widehat{V}_a$, Hwang et al. (2009) \cite{hwang_empirical_2009} and Dass et al. (2012) \cite{dass_confidence_2012} noted that assuming a common prior for $V_a$ for all areas $a$ could induce shrinkage in the resulting variance estimates and produce improved interval estimates for the parameters of interest. In this vein, Sugasawa et al. (2017)\cite{sugasawa_bayesian_2017} explore different priors for the sampling variances $V_a$, adopting a fully Bayesian approach to estimation. Alternatively, Polettini (2017) \cite{polettini_generalised_2017} induces shrinkage for the sampling variance estimates using a semiparametric Dirichlet process model with random variances.

\subsection{Variance Smoothing for Binary Response}

When response values are binary and the target of estimation is a small area proportion, it may be helpful to account for the mean-variance relationship observed in binary response data.  Generalized variance functions (GVFs), which model the functional relationship between the expectation and variance of a survey estimator, can be used as an alternative to linearization-based approximations or resampling methods for estimating $V_a$. If the model used is appropriate, the resulting modeled variance estimates could improve upon the direct variance estimates in terms of precision. An introduction to GVFs is provided in Chapter 7 of Wolter (2007) \cite{wolter_introduction_2007}. 

For small area estimation of proportions, several GVF-like approaches to variance estimation have been previously proposed based on treating the responses like binomial data. Liu et al. (2014) \cite{liu_hierarchical_2014} assume the following model for $V_a$:
\begin{equation}
V_a=\frac{p_a(1-p_a)}{n_a}DEFF_a
\end{equation}
where $DEFF_a$ denotes the design effect, defined as the ratio of the variance of $p_a$ under the implemented survey design to the the variance of $p_a$ under simple random sampling. As described in their paper, Liu et al. estimate design effects using available information on sample sizes and survey weights and treat them as known. Model-based estimates of $V_a$ can be produced by replacing the unknown $p_a$ values above with their direct estimators. Hawala and Lahiri (2018)  \cite{hawala_variance_2018} propose a similar GVF for count data. Maples (2016) \cite{maples_estimating_2016} similarly proposes a GVF for producing variance estimates based on estimating the design effect using additional information about any unequal weighting or clustering in the sampling procedure. Franco and Bell (2013) adopt a different strategy using a GVF to compute an effective sample size for each area of interest, which they use to fit a binomial model \cite{franco_applying_2013}.

Mohadjer et al. (2012) \cite{mohadjer_hierarchical_2012} similarly use a GVF to produce variance estimates for use in an area level model, assuming the following model
\begin{equation}
\log(V_a/p_a^2)=\eta_0+\eta_1\log(\widetilde{p}_a)+\eta_2\log(1-\widetilde{p}_a)+\eta_3\log(n_a)+\varepsilon_a
\end{equation}
where $\epsilon_a\sim N(0,\sigma_\epsilon^2)$ and $\widetilde{p}_a$ denotes a predictor of $p_a$ based on a model dependent solely on auxiliary covariate information and not explicitly on any direct estimates. 

The GVF approaches described thus far treat the resulting variance estimates as known, so the resulting Fay-Herriot estimates do not account for uncertainty in the variance model. Maples et al. (2009) \cite{maples_small-area_2009} address this by combining a GVF with a sampling model for the direct variance estimates. In particular, they assume Model (\ref{eqn:vsamp}) holds for the direct variance estimates $\widehat{V}_a$ and then propose the following linking model for $V_a$:
\begin{equation}
V_a\mid \alpha, \boldsymbol\gamma\sim IG(\alpha+1,\alpha \exp(\mathbf{z}_a^T\boldsymbol\gamma))
\end{equation}
where $\alpha$ controls the precision of the variance linking model, $\mathbf{z}_a$ are area level covariates and $\boldsymbol\gamma$ are corresponding coefficients estimated using an empirical Bayes approach. Maples et al. outline a procedure for using bootstrap sampling to estimate effective sample size for each area, which informs their choice for the degrees of freedom $d_a$ in the variance sampling model. They show that this model produces smoothed variance estimates that could help to correct underestimation by the direct variance estimators.

\subsection{Alternative sampling and linking models}

The linking and sampling mean models in the Fay-Herriot approach assume responses are continuous, but since $p_a$ are bounded between 0 and 1, it may be inappropriate to treat $\widehat{p}_a$ as Gaussian, especially when $p_a$ is close to 0 or 1 and when $V_a$ is large.  In the health and demography setting, Mercer et al. \cite{mercer_space-time_2015} apply a logit transformation to direct estimates of mortality rates before fitting a Fay-Herriot-type model. You and Rao (2002) \cite{liu_hierarchical_2014} proposed to use unmatched sampling and linking models, combining the sampling model given by Equation (\ref{eqn:msamp}) with an alternative linking model that transforms the finite population parameters of interest $p_a$ to make a Gaussian approximation more appropriate. As an example, Liu et al. (2014) \cite{liu_hierarchical_2014} considered the following logit-normal linking model:
\begin{equation}
\mathrm{logit}(p_a)\mid \boldsymbol\beta, \sigma_u^2\sim N(\mathbf{x}_a^T\boldsymbol\beta, \sigma_u^2)
\end{equation}
Mohadjer et al. (2012) \cite{mohadjer_hierarchical_2012} apply this model to estimation of adult literacy rates. Liu et al. also consider alternative models including a beta-logistic model combining a beta sampling model with the above logistic linking model, which accounts for the limited range of $\widehat{p}_a$ but will not reflect its true sampling distribution. Franco  and Bell (2013) \cite{franco_applying_2013}  and Chen et al. (2014)  \cite{chen_use_2014} consider binomial sampling models, treating observed area level counts as being drawn from a binomial distribution with size parameter given by some measure of effective sample size. As an alternative to unmatched sampling and linking models, Mercer et al. (2015) \cite{mercer_space-time_2015} describe an approach that uses Gaussian sampling and linking models to model $\mathrm{logit}(\widehat{p}_a)$ and $\mathrm{logit}(p_a)$.

\subsection{Spatial area level models}

The Fay-Herriot model has been extended to incorporate spatial and spatiotemporal random effects \cite{ghosh_generalized_1998, pratesi_small_2008}. These models leverage similarities between areas that are close in space or time, producing smoothed estimates and accounting for potential spatial or spatiotemporal structure in the response data. Chung et al. (2020) \cite{chung_bayesian_2020} noted that when informative area level covariate information is unavailable but responses are spatially correlated, using spatial random effects models may be especially effective.  In LMICs, covariate information is often limited and spatial random effects are often used in area level and unit level modeling. As an example, Mercer et al. (2015) use an area level model with spatiotemporal random effects to estimate child mortality rates in Tanzania \cite{mercer_space-time_2015}.

\section{Methods}\label{s:methods}

We assume that for all $a=1,\ldots, A$, we have direct estimates of area level proportions $\widehat{p}_a$ and corresponding variance estimates $\widehat{V}_a$. We propose a Bayesian joint model for the full data $(\widehat{\mathbf{p}}, \widehat{\mathbf{V}})$ that induces spatial smoothing for both the proportion and variance estimates. Our approach uses two sets of unmatched models, one for the estimated proportions $\widehat{\mathbf{p}}$ and one for the variance estimates $\widehat{\mathbf{V}}$, with these models being linked through the use of a generalized variance function. We use a spatial linking model for the proportions that induces spatial smoothing for both the proportions and the estimated variances.

\subsection{Mean model}

For modeling the direct estimates $\widehat{p}_a$, we use unmatched sampling and linking models, combining a Gaussian sampling model with a spatial logit-Gaussian linking model:
\begin{equation}\label{e:mean-sampling}
\widehat{p}_a\mid p_a, V_a \stackrel{ind}{\sim} N(p_a, V_a).
\end{equation}
\begin{equation}\label{e:mean-linking}
\mathrm{logit}(\mathbf{p})\mid \boldsymbol\beta, \sigma_u^2, \phi \sim \boldsymbol{\mathcal{N}}(\mathbf{X}\boldsymbol\beta, \boldsymbol\Sigma_{BYM2}( \sigma_u^2, \phi)).
\end{equation}
In the above, $p_a$ denotes the finite population area-specific proportion and $V_a$ denotes the sampling variance of the direct estimator $\widehat{p}_a$. We use the shorthand $\mathrm{logit}(\mathbf{p})$ to denote the vector $(\mathrm{logit}(p_1),\ldots, \mathrm{logit}(p_A))^T$, which we assume is drawn from a multivariate Gaussian distribution with mean $\mathbf{X}\boldsymbol\beta$, where $\mathbf{X}$ is an $A\times (p+1)$ design matrix containing covariate information and $\boldsymbol\beta$ is a $(p+1)$-vector containing the intercept and corresponding coefficients. Finally $\boldsymbol\Sigma_{BYM2}( \sigma_u^2, \phi)$ denotes a spatial covariance matrix dependent on marginal variance parameter $\sigma_u^2$ and spatial correlation parameter $\phi$. We use the BYM2 model, a reparametrization of the Besag-York-Molli\'e \cite{besag_bayesian_1991} model proposed by Riebler et al. (2016) \cite{riebler_intuitive_2016} which determines the structure of $\boldsymbol\Sigma_{BYM2}( \sigma_u^2, \phi)$. Below, we review the BYM2 model, rewriting the mean linking model as follows for clarity:
\begin{equation}
\mathrm{logit}(\mathbf{p})= \mathbf{X}\boldsymbol\beta+\mathbf{u}
\end{equation}
\begin{equation}
\mathbf{u}\mid \boldsymbol\sigma_u^2, \phi \sim \boldsymbol{\mathcal{N}}(\mathbf{0}, \boldsymbol\Sigma_{BYM2}( \sigma_u^2, \phi)).
\end{equation}
Under the BYM2 model, we assume $\mathbf{u}$ can be partitioned into an unstructured component $\widetilde{\mathbf{u}}_1$ and a structured spatial component $\widetilde{\mathbf{u}}_{2*}$:
\begin{equation}
\mathbf{u}=\sigma_u\left(\sqrt{1-\phi}\widetilde{\mathbf{u}}_1+\sqrt{\phi}\widetilde{\mathbf{u}}_{2*}\right)
\end{equation}
We assume $\widetilde{\mathbf{u}}_1\sim N(\mathbf{0}, \mathbf{I})$ is a vector of iid Gaussian random area effects and assume an intrinsic conditional autoregressive (ICAR) Gaussian prior for $\widetilde{\mathbf{u}}_{2*}$. The ICAR prior, as proposed by Besag et al. (1991) \cite{besag_bayesian_1991}, assumes that spatial components $\widetilde{u}_{2a*}$ and $\widetilde{u}_{2b*}$, representing the values of $\widetilde{\mathbf{u}}_{2*}$ for areas $a$ and $b$, are correlated if areas $a$ and $b$ are defined to be neighbors. Under an ICAR prior, we assume that for a particular area $a$, the mean of $\widetilde{u}_{2a*}$ is equal to the mean of all neighboring effects and the precision of $\widetilde{u}_{2a*}$ is proportional to the number of neighbors. Using this parameterization, $\sigma_u$ denotes the marginal variance of $\mathbf{u}$ and $\phi$ represents the proportion of variation assigned to the spatial component.

Under a BYM2 model, $\mathbf{u}$ has the covariance matrix
\begin{equation}
\boldsymbol\Sigma_{BYM2}( \sigma_u^2, \phi)=\sigma_u((1-\phi)\mathbf{I}+\phi \mathbf{Q}_{*}^{-})
\end{equation}
Here,  $\mathbf{Q}_{*}$ denotes the precision matrix of $\widetilde{\mathbf{u}}_{2*}$ and $\mathbf{Q}_{*}^{-}$ is its generalized inverse. Note that the precision matrix implied by the ICAR prior, $\mathbf{Q}_{*}$, is singular,  yielding an improper prior. To ensure identifiability, we must place a sum-to-zero constraint on $\mathbf{u}$. In order to make the marginal variance parameter $\sigma_u$ interpretable, we scale $\mathbf{Q}_{*}$ to make the geometric mean of the marginal variances equal to one, as recommended by Riebler et al. (2016) \cite{riebler_intuitive_2016} 

\subsection{Variance model}

We similarly use unmatched models for the corresponding variance estimates $\widehat{\mathbf{V}}$, using a chi-squared sampling model with a log-normal linking model. We use the chi-squared sampling model described in Equation (\ref{eqn:vsamp}), assuming that for all $a$, the variance estimate $\widehat{V}_a$ is an unbiased estimator of $V_a$. The linking model assumes the true $\log(V_a)$ values are Gaussian distributed with expected values given by a generalized variance function $f(p_a, \mathbf{z}_a; \boldsymbol\gamma)$ whose inputs are the area proportion $p_a$, other area level predictors $\mathbf{z}_a$, and parameters $\boldsymbol\gamma$. We can write down the unmatched models as follows:
\begin{equation}
\frac{d_a\widehat{V}_a}{V_a}\mid d_a, V_a\stackrel{ind}{\sim}\chi^2_{d_a}
\end{equation}
\begin{equation}\label{e:var-linking}
\log(V_a)\mid p_a, \mathbf{z}_a, \boldsymbol\gamma, \sigma_\tau^2\stackrel{ind}{\sim} N\left(f(p_a, \mathbf{z}_a; \boldsymbol\gamma), \sigma_\tau^2\right)
\end{equation}
Here, $d_a$ denotes the degrees of freedom parameter for area $a$, which we determine based on the survey design and sample size as discussed below. We use $\sigma_\tau^2$ to denote the variance of the linking model errors which allow for area-specific deviations from the linking model.

We define the generalized variance function as follows:
\begin{equation}
f(p_a, \mathbf{z}_a; \boldsymbol\gamma)=\gamma_0+\gamma_1\log(p_a(1-p_a))+\gamma_2\log(n_a)
\end{equation}
where $n_a$ denotes the sample size for area $a$. Note that if we set $\gamma_0=0, \gamma_1=1, \gamma_2=-1$, the right hand side resembles the logarithm of the binomial variance. As such, this GVF can be viewed as a generalized version of the binomial variance. The GVF used here could also be altered to introduce additional covariates 
or different functional relationships between $p_a$ and $V_a$. We can view the variance linking model (\ref{e:var-linking}) as a prior that shrinks the estimate $\widehat{V}_a$ towards a model-based prediction dependent on the binomial mean-variance relationship.

As described above, the mean linking model induces spatial smoothing for estimates $\widehat{p}_a$. By combining the mean and variance models and incorporating the means $p_a$ into the GVF, we induce spatial correlation into the resulting samples of $V_a$, potentially aiding estimation in areas with fewer samples.

We treat the degrees of freedom parameter $d_a$ as known for all areas $a$. The appropriate choice for $d_a$ depends on the sampling design. As mentioned above, if the data for a given area were iid Gaussian (for example, reflecting simple random sampling with replacement), the typical variance estimator would follow a $\chi^2$ distribution with $d_a=n_a-1$ degrees of freedom. However, for sampling without replacement and cluster sampling designs, other choices of $d_a$ may be more appropriate depending on how $\widehat{V}_a$ is computed for each area. Maples et al. (2009) \cite{maples_small-area_2009} outline a resampling procedure for estimating degrees of freedom for their variance sampling model. When computing variance estimates from DHS data, we use a simplified variance estimator based on the with-replacement variance estimator for multistage designs presented in Equation (4.6.2) of S\"arndal et al. (1993)\cite{sarndal_model_2003}, which is computed as a sum over clusters:
\begin{equation}
\widehat{V}_a=\frac{1}{m_a(m_a-1)}\sum_{j\in S_{1a}}\left(\frac{\widehat{t}_{aj}}{\pi_j}-\widehat{t}_a\right)^2
\end{equation}
where $S_{1a}$ denotes the set of indices of sampled clusters for area $a$, $m_a$ denotes the number of sampled clusters, and $\pi_j$ denotes the probability of sampling cluster $j$. Finally $\widehat{t}_{aj}$ denotes the direct estimator for the total for cluster $j$ in area $a$ and $\widehat{t}_a$ denotes the direct estimator for the total of area $a$. Since this is a sum of squared error terms over $m_a$ clusters, we set $d_a$ to be equal to $m_a-1$.

\subsection{Estimation}
We adopt a fully Bayesian approach to estimation by placing priors on the following hyperparameters:
\begin{equation}
\{\boldsymbol\beta, \sigma_u^2, \phi, \boldsymbol\gamma,\sigma_\tau^2\}\sim \Pi(\boldsymbol\theta)
\end{equation}
where $\boldsymbol\theta$ denotes any parameters used to specify the priors. Details on the priors used in each example are provided in Appendix \ref{a:est}. We compute approximate posterior distributions for $p_a$ for all areas $a$ using Markov chain Monte Carlo sampling as implemented in the Stan programming language \cite{carpenter_stan_2017}.  Functions for fitting the models described above have been collected in an R package called \texttt{VSALM} available at \texttt{https://github.com/peteragao/VSALM} and the code for the below simulations and analysis is available at the associated repository \texttt{https://github.com/peteragao/VSALM-paper}. In this context, the Bayesian approach offers a number of potential benefits. In particular, we are able to sample from the joint posterior distributions for the proportions of interest for all areas, giving a natural way to quantify uncertainty and also enabling comparisons between areas. Moreover, the sampling approach implemented in Stan is fast and flexible, enabling users to fit and compare potential models quickly.

\section{Simulations}\label{s:sim}

\subsection{Population generating model}

We use simulations to evaluate our spatial variance smoothing estimator, comparing its performance with that of the direct weighted H\'ajek estimator and an estimator derived from a model without variance smoothing. For our simulations, we generate an artificial population that mimics data from the 2018 Nigeria DHS. First,  we generate synthetic cluster locations across Nigeria using a pixel grid of estimated population counts for Nigeria in 2006 (mimicking the sampling frame used for the DHS survey) \cite{worldpop_global_2006}. For each of the 73 strata used for the DHS, we sample 300 pixels without replacement with probability proportional to population. These sampled pixels represent enumeration areas or clusters. For each cluster location, we randomly generate cluster sizes $N_c\sim \mathrm{Poisson(10)}$, yielding a population of $N=\sum\limits_c N_c$ individuals.

For each individual $i$ in our population, we  generate data using a population generating model motivated by the models used by Corral et al. (2020) \cite{corral_pull_2020} (see Section 7.2) and Gao and Wakefield (2022) \cite{gao_smoothed_2022}. For each cluster $c$, we simulate cluster level covariate information as follows:
\begin{enumerate}
	\item The covariate $x_{1,c}$ is the realized value of a binary random variable $X_{1,c}$ with $P(X_{1,c} = 1) = 0.5$;
	\item The covariate $x_{2,c}$ is the realized value of a binary random variable $X_{2,c}$ with $P(X_{2,c} = 1) = 0.3 + 0.5\frac{a(c)}{37}$, where $a(c)$ is the index of the area containing cluster $c$;
	\item The covariate $x_{3,c}=x_{3,a(c)}$ is obtained from a  $37\times 1$ ICAR random vector with marginal variance 1 for the Admin-1 areas.
	\item The covariate $x_{4,c}=x_{4,a(c)}$ is obtained from a $774\times 1$ ICAR random vector with marginal variance 1 for the Admin-2 areas.
	\item The covariate $x_{5,c}$ is obtained from a random vector generated using a stochastic partial differential equation (SPDE) -based approximation \cite{lindgren_explicit_2011} to a Gaussian process with Mat\'ern covariance with smoothness parameter 1 and marginal variance of 1.
\end{enumerate} 
Maps of these covariates are provided in Appendix \ref{a:cov-maps}. Based on these covariates, we simulate a cluster level risk parameter $q_c$ for each cluster from the following model:
\begin{multline}
\mathrm{logit}(q_c)=\mathrm{logit}(\mu)+ 0.25x_{1,c} - 0.25x_{2,c} + 0.5x_{3,c} + 0.25x_{4,c} + 0.25x_{5,c} +u_{a(i)}+v_{c}
\end{multline}
where $u_{a(i)}\stackrel{iid}{\sim} N(0,0.25^2)$ are independent and identically distributed area level random effects, and $v_c\stackrel{iid}{\sim} N(0,0.5^2)$ represents random and independent and identically distributed cluster level effects. In the above, $\mu$ denotes the global superpopulation prevalence. The covariates are held constant for all simulations, but the response variables and random area and cluster effects are resampled for each new simulation. We repeatedly generate $Y_i\mid q_{c(i)}\sim\mathrm{Bernoulli}(q_{c(i)})$,  where $c(i)$ is the cluster of individual $i$. For each simulation, we can thus compute true population Admin-1 area level proportions $p_a$. 
 
To obtain our simulated samples, we use a cluster sampling design. In each simulation, we sample eight clusters from each stratum, keeping all individuals in sampled clusters.  We compute sampling weights $w_i$ for each individual from the corresponding inverse inclusion probabilities.

We compare our unmatched joint smoothing model-based estimator with the direct \textbf{H\'ajek} estimator and a number of alternative model-based estimates. First, we consider spatial joint sampling (\textbf{Spatial Unmatched JS}) and non-spatial joint sampling (\textbf{Unmatched JS}) models, where the non-spatial version is obtained by replacing the BYM2 prior for the area effects $\mathbf{u}$ with an iid multivariate Gaussian prior. We also consider an estimator produced using a model that omits the variance smoothing model entirely, which we refer to as the \textbf{mean smoothing (MS)} model-based estimator. This model is specified using the unmatched models (\ref{e:mean-sampling}) and (\ref{e:mean-linking}) but treating $\widehat{V}_a$ as known for all $a$. We consider both spatial (\textbf{Spatial Unmatched MS}) and non-spatial versions (\textbf{Unmatched MS}). For all model-based estimators, we adopt a fully Bayesian approach to inference as described in Section \ref{s:methods}. We obtain point estimates $\widehat{p}_a$ and 90\% credible interval estimates $(l_a, u_a)$ by sampling from these approximate posterior distributions. Further details on the estimation procedure are provided in the Appendix.

For each simulation, we calculate several performance metrics, including root mean squared error (RMSE) and mean absolute error (MAE). We also compute the average coverage of the 90\% interval estimates and the mean interval lengths (MIL) across all areas. For a single simulation, these metrics are defined as follows.
\begin{align}
\mathrm{RMSE}(\widehat{p}_a)&=\sqrt{\frac{1}{A}\sum_{a}(p_a-\widehat{p}_a)^2}\label{e:rmse}\\
\mathrm{MAE}(\widehat{p}_a)&=\frac{1}{A}\sum_{a}|p_a-\widehat{p}_a|\\
\mathrm{Cov}_{90}(\widehat{p}_a)&=\frac{1}{A}\sum_{a}\mathbf{1}\{p_a\in(l_a, u_a)\}\\
\mathrm{MIL}_{90}(\widehat{p}_a)&=\frac{1}{A}\sum_{a}(u_a - l_a)\label{e:mil}
\end{align}

\begin{table}
	\centering
	\scriptsize
	
\begin{tabular}{l|rrrr|rrrr}
\toprule
&\multicolumn{4}{c|}{$\mu = 0.1$}&\multicolumn{4}{c}{$\mu = 0.5$}\\
\textbf{Method} & \textbf{RMSE} &  \textbf{MAE}& \textbf{90\% Cov.} & \textbf{MIL} &\textbf{RMSE} &  \textbf{MAE}& \textbf{90\% Cov.} & \textbf{MIL}\\
\midrule
Direct (Hájek) & 3.44 & 2.63 & 83 & 10.42 & 6.16 & 4.92 & 85 & 19.55\\
\midrule
Unmatched MS & 3.53 & 2.55 & 82 & 9.10 & 5.80 & 4.62 & 85 & 17.48\\
Spatial Unmatched MS & 3.28 & 2.39 & 83 & 8.60 & 5.61 & 4.46 & 85 & 17.20\\
\midrule
Unmatched JS & 3.24 & 2.46 & 89 & 9.78 & 6.44 & 5.06 & 89 & 20.20\\
Spatial Unmatched JS & 3.03 & 2.28 & 90 & 9.34 & 6.19 & 4.87 & 90 & 19.72\\
\bottomrule
\end{tabular}

	\caption{RMSE ($\times 100$), MAE ($\times 100$), coverage rates, and mean interval length ($\times 100$) of estimators of area level means across 1,000 simulated populations with spatially correlated binary responses based on sample data obtained via informative sampling. The reduced model omits one of the spatial covariates in the full model.}
	\label{tab:sp-inf-model}
\end{table}

We consider two sets of simulations with differing global prevalence rates.  We let $\mu=0.1$ for the first set, which is similar to the overall HIV positivity rate in the Malawi data. For the second set, we let $\mu=0.5$,  which is similar to the national MCV-1 vaccination rate in the Nigeria data. Table \ref{tab:sp-inf-model} summarizes results for our two sets of simulations. In each setting, the results represent the average values of the metrics (\ref{e:rmse})-(\ref{e:mil})  across 1,000 simulated populations. We observe that in the low prevalence examples, the spatial unmatched joint model-based estimates perform best in terms of RMSE and MAE. Moreover, prediction intervals constructed based on the direct estimator and the mean-only smoothing model-based estimators tend to exhibit undercoverage, whereas the joint model-based intervals achieve closer to nominal coverage. In the moderate prevalence examples, the joint model-based estimates perform slightly worse than the mean-only model-based estimates in terms of the RMSE and MAE; however, the H\'ajek and mean-only model intervals show slight undercoverage. The joint modeling approach thus yields slightly more conservative prediction intervals which may be desirable for decision making.

\section{Applications}\label{s:app}

We apply our joint smoothing model-based estimator to two examples involving DHS data, demonstrating its use for a low prevalence indicator (Malawi HIV prevalence rates) and for a moderate prevalence indicator (Nigeria measles vaccination rates). We show that our method induces spatial smoothing of estimated variances and produces more conservative interval estimates than an approach using an area level model that only smooths means. 

For both examples, we compare direct weighted estimation with the model-based smoothing methods described above. We first fit both the spatial mean smoothing unmatched model (\textbf{Spatial Unmatched MS}), which omits the variance model, as well as the  full spatial joint smoothing unmatched model (\textbf{Spatial Unmatched JS}) For all models, we use no covariates and we compute approximate posterior distributions for all area level proportions $p_a$ and obtain corresponding point and interval estimates by sampling from these posteriors. 
\begin{figure}
	\centering
	\includegraphics[scale=.45]{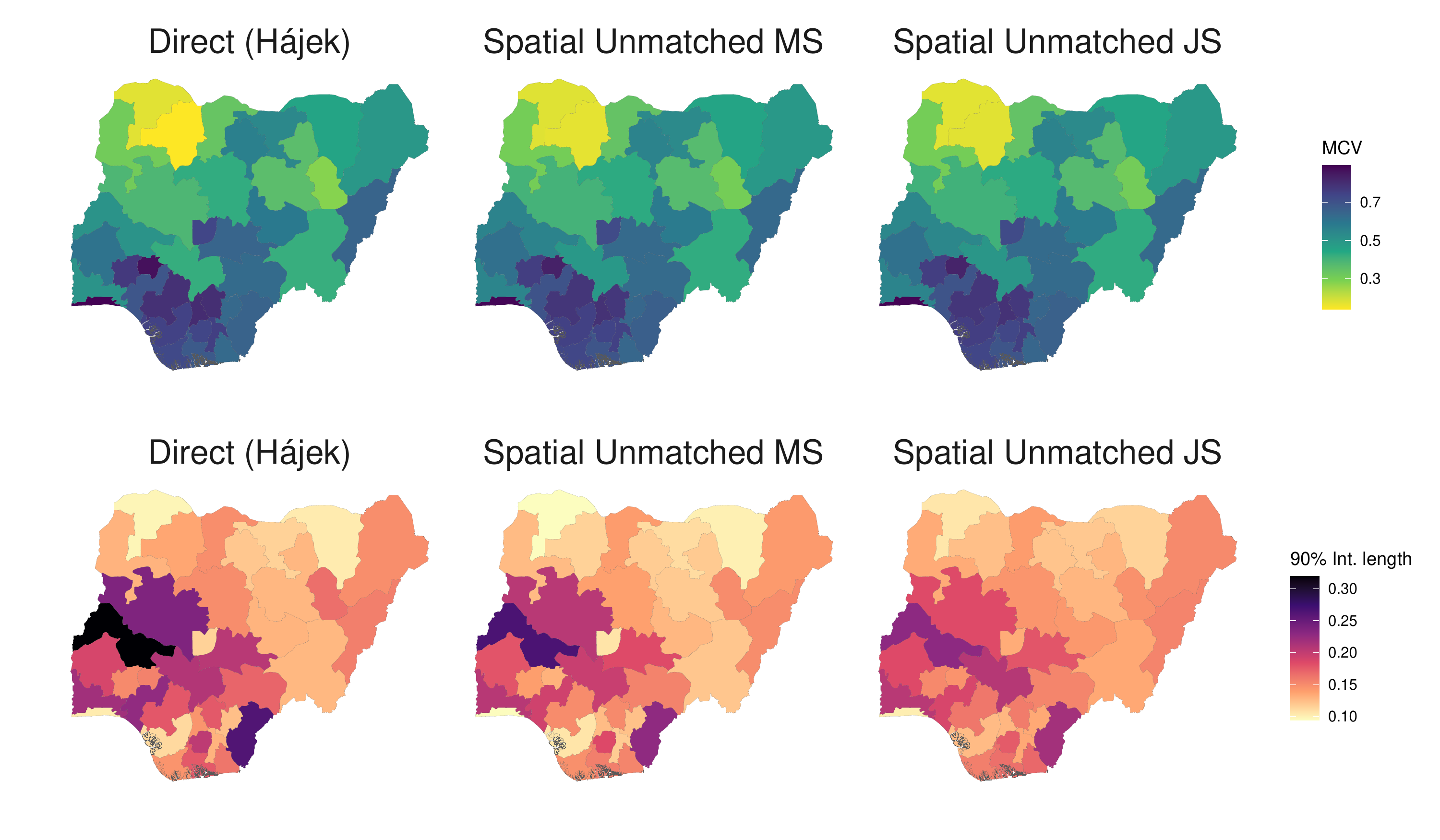}
	\caption{Direct and model-based point estimates (top) and  length of corresponding 90\% interval estimates (bottom) of vaccination coverage rate for first dose of measles-containing-vaccine (MCV1) among children aged 12–23 months in Nigeria, 2018.}
	\label{fig:map-nga-mcv}
\end{figure}

\begin{figure}
	\centering
	\includegraphics[scale=.75]{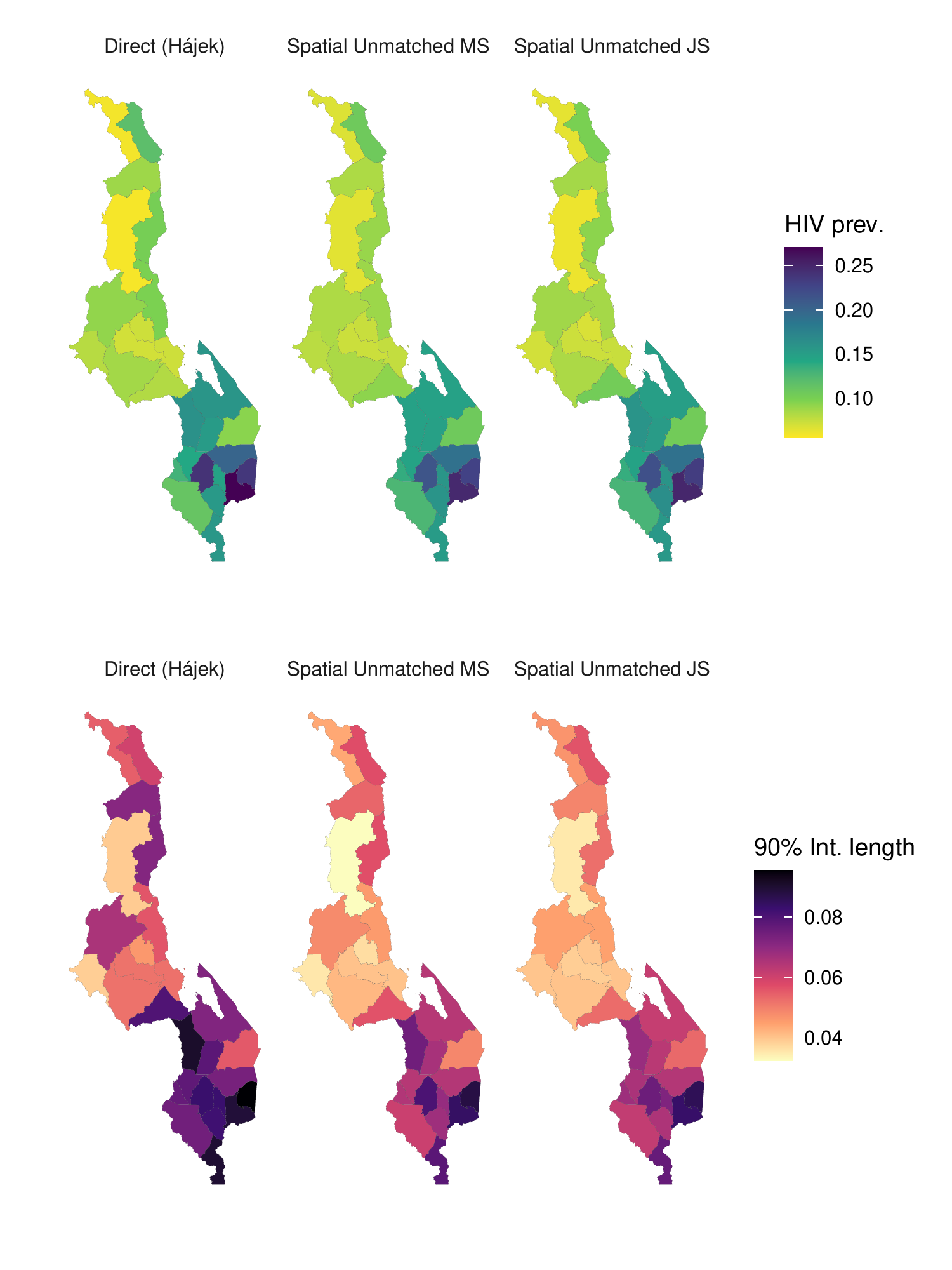}
	\caption{Direct and model-based point estimates (top) and  length of corresponding 90\% interval estimates (bottom) of HIV prevalence rate for women aged 15-49 in Malawi, 2015-2016.}
	\label{fig:map-mwi-hiv}
\end{figure}

Figure \ref{fig:map-nga-mcv} compares point estimates of MCV-1 coverage rates (top) and the length of interval estimates (bottom) for Admin-1 areas among children aged 12-23 months in Nigeria in 2018.  Figure \ref{fig:map-mwi-hiv} similarly provides point estimates of HIV prevalence rates (top) and the length of interval estimates (bottom) for Admin-1 areas for women aged 15-49 in Malawi, 2015-2016. For both examples, the bottom set of maps illustrates the estimated uncertainty of the direct and model-base estimates using the length of 90\% credible intervals. In general, we observe that the point estimates agree well for all three methods. However, we observe some spatial smoothing of the interval lengths, suggesting that the joint smoothing model induces spatial smoothing of the direct estimator variances $V_a$.

\begin{figure}
	\centering
	\includegraphics[scale=.6]{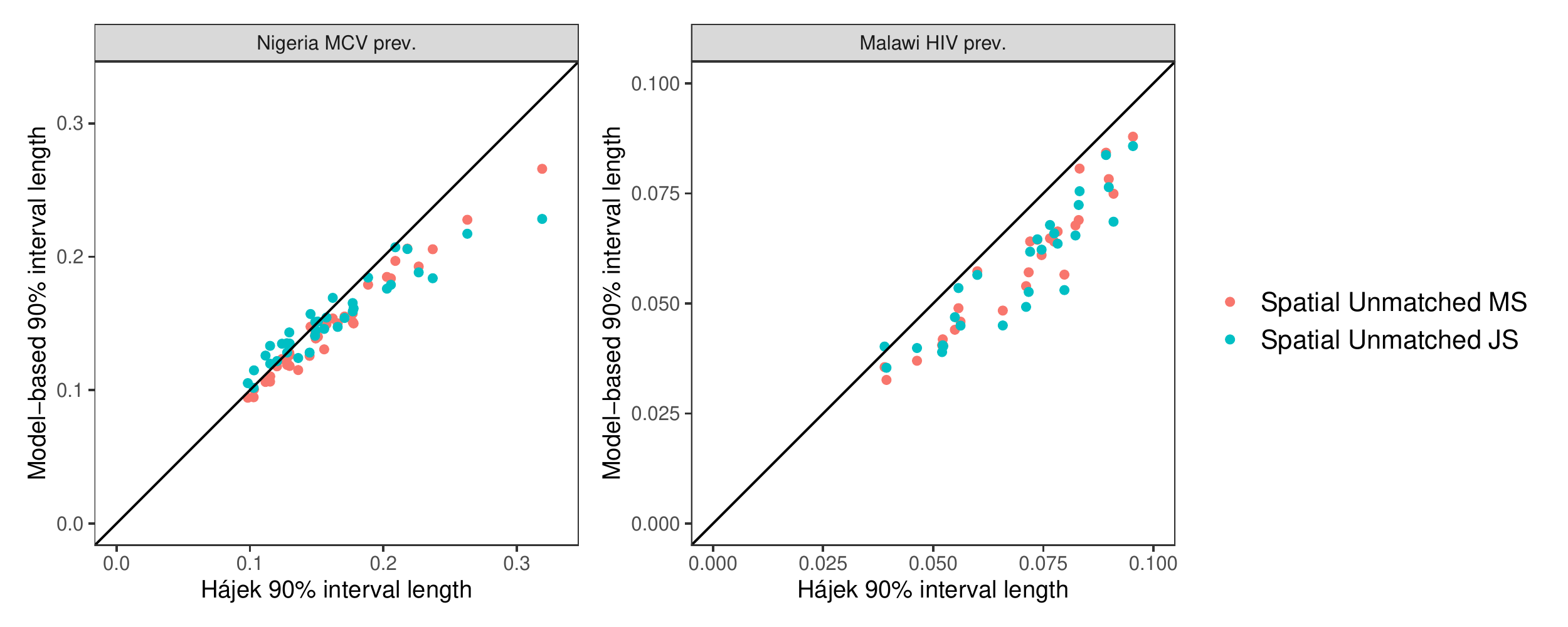}
	\caption{Comparison of model-based 90\% credible interval lengths with H\'ajek 90\% confidence interval lengths for Malawi HIV example (left) and Nigeria MCV example (right).}
	\label{fig:int-length-scatter}
\end{figure}

Figure \ref{fig:int-length-scatter} provides a scatter plot comparing the model-based 90\% credible interval lengths produced by the mean smoothing and joint smoothing models with the design-based 90\% confidence interval length associated with the H\'ajek estimator. Here, we see that for the Malawi example (left), the credible interval lengths are similar for the mean smoothing and joint smoothing models. However, for the Nigeria example (right), the joint smoothing intervals are more conservative when the H\'ajek intervals are short and narrower when the H\'ajek intervals are long.  In addition to the spatial smoothing of interval lengths seen in Figures \ref{fig:map-nga-mcv} and \ref{fig:map-mwi-hiv}, this suggests the joint model can help smooth variance estimates globally. 

\section{Discussion}\label{s:dis}

Our proposed model-based estimator for small area proportions combines several proposed extensions of the basic Fay-Herriot area level model, including using unmatched linking and sampling models to address the non-Gaussian response data \cite{you_small_2002}, incorporating spatial smoothing via correlated random effects in the mean linking model \cite{pratesi_small_2008}, and introducing a variance smoothing model so that the resulting estimators exhibit both smoothed means and variances \cite{you_small_2006, maiti_prediction_2014, sugasawa_bayesian_2017}. We propose a spatial joint smoothing model and adopt a fully Bayesian approach to estimation, which facilitates quick computation of point and interval estimates. Through simulation and application, we have shown that inferences based on our model can improve upon those based on a model that only incorporates smoothing of means. Interval estimates obtained from our model can correct for the undercoverage seen in models that only smooth means, suggesting our model may more accurately account for uncertainty in estimated variances of direct weighted estimators.

For our clustered binary response data, the variance smoothing model we have adopted may help address undercoverage of interval estimates caused by treating variances of direct weighted estimators as known. However, for other designs and contexts, such a model may be inappropriate. In general, the choice of variance sampling and linking models should depend on a number of factors including any clustering and stratification in the design as well as the distribution and presumed mean-variance relationship of the response variables.

Moreover, we acknowledge that our variance smoothing model is a simplification of the true distribution of design-based variances $V_a$.  in particular, the use of a chi-squared distribution for the variance sampling model relies upon the assumption that the direct estimator of variance $\widehat{V}_a$ for a particular area $a$ is computed as the sum of several squared Gaussian terms. Since our data is non-Gaussian, this assumption may be violated and other sampling models for $\widehat{V}_a$, such as a Gaussian model, could be explored i future work. Moreover, within each area, we have assumed that the variance for each stratum is equal, but this assumption may be inappropriate for our data since each area of interest is divided into urban and rural subregions, which may be qualitatively different from one another. Finally, the appropriate number of degrees of freedom $d_a$ depends on the specific design used; using resampling methods like those explored by Maples et al. (2009) \cite{maples_small-area_2009} to choose $d_a$ may help improve the fit of the variance sampling model. 

Although we have presented one approach applied to two different types of problems, in practice, for decision making purposes, different estimation problems may have varying priorities. For example, when implementing targeted vaccination programs, it is important to identify communities with especially low vaccination rates, whereas designing policy for providing resources associated with HIV involves identifying communities with high rates of positivity. Given that the variance of a direct prevalence rate estimator may depend on its expected value,  various modeling decisions such as choosing to apply a transformation for $\widehat{p}_a$ may lead to different results depending on the expected value of $\widehat{p}_a$. As such, it is crucial to carefully consider the distribution of direct estimators before selecting a model. Above, we have used unmatched sampling and linking models for the area level proportions, but we also considered first computing the logit-transformed direct estimators and then applying matched sampling and linking models treating both $\widehat{p}_a$ and $p_a$ as Gaussian random variables. In our simulations and application, this approach did not outperform the unmatched models we adopted, but future research could help illustrate when such an approach could be useful.
 
When mapping subnational health and demographic indicators in LMICs, unit level models, and in particular geostatistical models using spatial Gaussian processes, are often used as they allow estimates to be generated at arbitrary resolutions and can incorporate unit level covariate information. However, such approaches may often struggle to account for design effects such as those caused by clustering and informative sampling. While unit level models may be able to generate prevalence estimates at the individual cluster level, aggregating those cluster level estimates upwards to produce area level estimates may introduce additional errors and lead to improperly calibrated interval estimates \cite{fuglstad_two_2022, paige_design-_2022, paige_spatial_2022}. Area level models are specified to generate estimates for a preselected set of regions.  Moreover, area level models are often simpler and faster to implement than unit level models. For these reasons, we have explored the feasibility of using area level models to generate maps of health indicators such as vaccination rates and disease prevalence rates in LMICs. Our method, like many area level methods, directly accounts for survey design by incorporating available sampling weight information. By incorporating a spatial variance smoothing model and using unmatched sampling and linking models, we are able to address some of the difficulties related to applying area level models for use in this specific context.

\bibliographystyle{WileyNJD-AMA}
\bibliography{./vsfh}

\appendix
\section{Parameter estimation\label{a:est}}

Below we provide further details on the estimation process and priors used for each method described in the simulations and applications.

\subsection{Direct estimation}

We use the R package \texttt{survey} for computing direct weighted H\'ajek estimators (and corresponding variance estimates) for all areas.

\subsection{Mean-smoothing model-based estimation}

We adopt a fully Bayesian approach to estimating the mean-smoothing unmatched model described above, assuming priors for model parameters and then using MCMC as implemented in the R package \texttt{STAN} to sample from the posterior distributions of the area level proportions $p_a$ for all $a=1,\ldots, A$. We place a $N(0, 1000)$ prior on the area level model intercept and fixed effects.  We use penalized complexity priors for the variance parameter $\sigma_u$, as described by Simpson et al. (2017) \cite{simpson_penalising_2017} We specify these priors such that $P(\sigma_u>1)=0.01$. For the spatial models, we place a $\mathrm{Beta}(1/2,1/2)$ prior on the spatial correlation prior $\phi$.

\subsection{Joint-smoothing model-based estimation}
We keep the same priors as for the mean-smoothing model on area level model intercept and fixed effects as well as the variance parameters $\sigma_u$ and for the spatial models, $\phi$. We then place a penalized complexity prior for the variance parameter $\sigma_\tau$ such that $P(\sigma_\tau>1)=0.01$. We place $N(0,1)$ prior on $\gamma_0$ and $N(1,.5)$ prior on $\gamma_1$ and  $N(-1,.5)$ prior on $\gamma_2$ to shrink the resulting variances estimates towards that of a binomial random variable.

\section{Additional results}

\subsection{Covariate maps\label{a:cov-maps}}

\begin{figure}
	\centering
	\includegraphics[scale=.6]{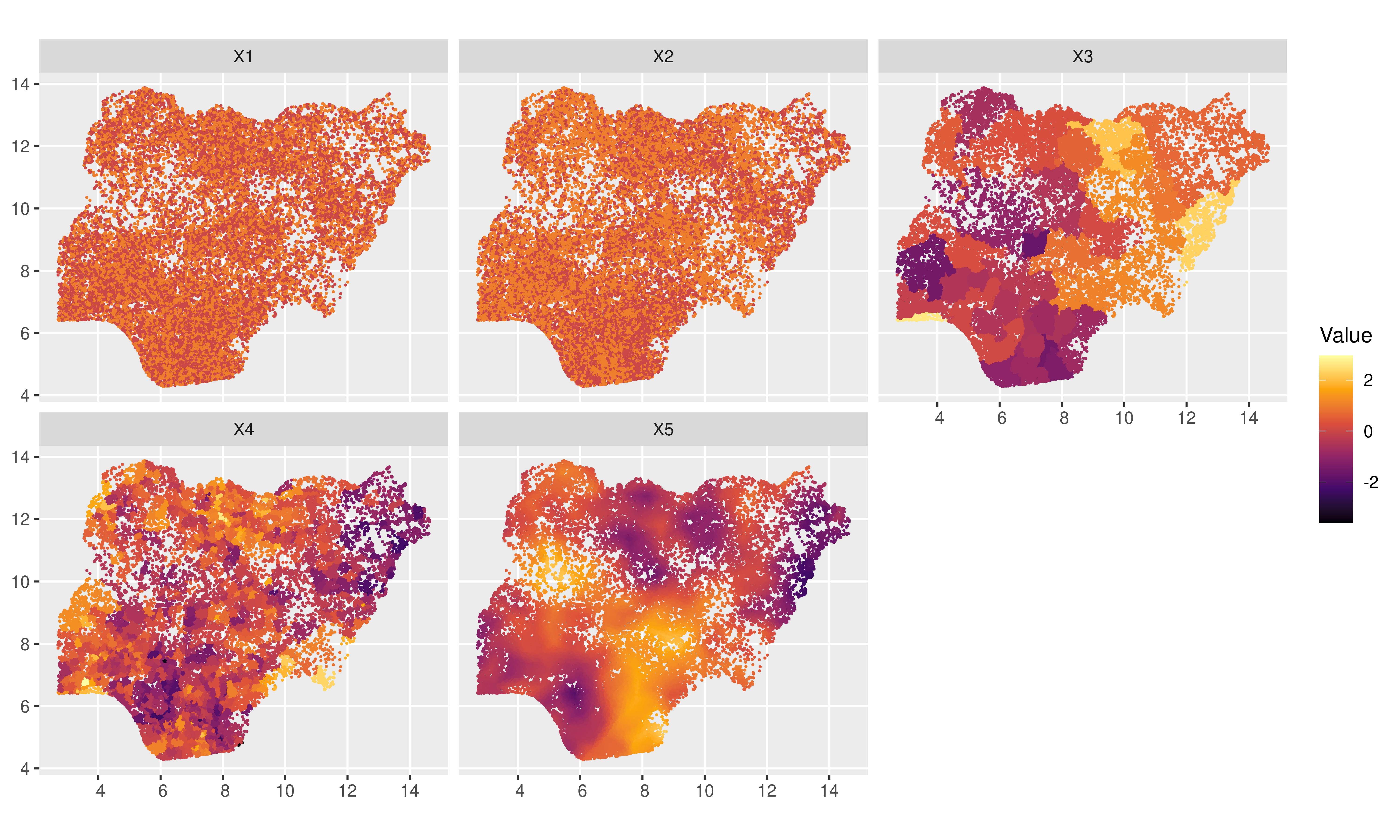}
	\caption{Simulated cluster locations and covariate values used to generate population data.}
	\label{fig:cov-maps}
\end{figure}

Figure \ref{fig:cov-maps} provides maps of the simulated cluster locations and covariate values used in the simulations described in Section \ref{s:sim}.

\subsection{Large sample simulations}

Table \ref{tab:lss} provides results for an additional set of simulations that were identical to the simulations with $\mu=0.5$ described in Section \ref{s:sim}, except with a larger sample size, with twenty-five clusters sampled per stratum rather than the eight used in the main text. The results illustrate that for large sample sizes, the joint smoothing and mean smoothing model-based estimators perform similarly to the direct weighted estimators with 90\% prediction interval coverage rates that are close to nominal.

\begin{table}
	\centering
	\scriptsize
	
\begin{tabular}{lrrrr}
	\toprule
	\textbf{Method} & \textbf{RMSE} &  \textbf{MAE}& \textbf{90\% Cov.} & \textbf{MIL}\\
	\midrule
	Direct (Hájek) & 2.40 & 1.91 & 90 & 7.98\\
	\midrule
	Unmatched MS & 2.35 & 1.87 & 90 & 7.87\\
	Spatial Unmatched MS & 2.33 & 1.86 & 90 & 7.80\\
	\midrule
	Unmatched JS & 2.37 & 1.89 & 91 & 7.99\\
	Spatial Unmatched JS & 2.35 & 1.87 & 91 & 7.92\\
	\bottomrule
\end{tabular}

	\caption{RMSE ($\times 100$), MAE ($\times 100$), coverage rates, and mean interval length ($\times 100$) of estimators of area level means across 1,000 simulated populations with spatially correlated binary responses based on sample data obtained via informative sampling. The reduced model omits one of the spatial covariates in the full model.}
	\label{tab:lss}
\end{table}

\subsection{Applications}

Tables \ref{tab:mcv-param} and \ref{tab:hiv-param} provide point estimates and corresponding 90\% prediction intervals for model hyperparameters.

Tables \ref{tab:mcv-detail} and \ref{tab:hiv-detail} provide full estimates with prediction intervals for all areas of interest for the methods described in the manuscript applied to both the measles vaccination rate and HIV prevalence rate applications. 

\begin{table}
	\centering
	
\begin{tabular}{r|rl|rl}
\toprule
	Parameter & \multicolumn{2}{|c}{\textit{Spatial Unmatched MS}}&\multicolumn{2}{|c}{\textit{Spatial Unmatched JS}}\\
\midrule
$\mathrm{logit}(\mu)$&0.36 & (0.22, 0.5) & 0.36 & (0.23, 0.49)\\
$\sigma_u$&0.71 & (0.54, 0.96) & 0.71 & (0.54, 0.94)\\
$\phi$&0.84 & (0.38, 1) & 0.86 & (0.39, 1)\\
$\gamma_0$&&  &-0.73 & (-2.22, 0.74)\\
 $\gamma_1$& &  & 0.47 & (-0.14, 1.01)\\
 $\gamma_2$ &&  &-0.9 & (-1.18, -0.63)\\
 $\sigma_\tau$& &  &  0.42 & (0.29, 0.59)\\
\bottomrule
\end{tabular}
	\caption{Point estimates and 90\% interval estimates for model parameters for Nigeria measles vaccination example}
	\label{tab:mcv-param}
\end{table}

\begin{table}
	\centering
	
\begin{tabular}{r|rl|rl}
	\toprule
	Parameter & \multicolumn{2}{|c}{\textit{Spatial Unmatched MS}}&\multicolumn{2}{|c}{\textit{Spatial Unmatched JS}}\\
\midrule
$\mathrm{logit}(\mu)$&-2.03 & (-2.14, -1.93) & -2.02 & (-2.14, -1.92)\\
$\sigma_u$&0.41 & (0.28, 0.62) & 0.43 & (0.3, 0.64)\\
$\phi$&0.89 & (0.36, 1) & 0.88 & (0.36, 1)\\
$\gamma_0$& &  &  -0.22 & (-1.9, 1.5)\\
$\gamma_1$& &  &0.91 & (0.51, 1.27)\\
$\gamma_2$& &  & -0.96 & (-1.29, -0.64)\\
 $\sigma_\tau$&&  & 0.19 & (0.07, 0.37)\\
\bottomrule
\end{tabular}

	\caption{Point estimates and 90\% interval estimates for model parameters for Malawi HIV prevalence example}
	\label{tab:hiv-param}
\end{table}

 	\begin{table}
 		\centering
 		\tiny
 		
\begin{tabular}{l|rl|rl|rl}
	\toprule
	State& \multicolumn{2}{c|}{\textit{H\'ajek}} &  \multicolumn{2}{c|}{\textit{Spatial Unmatched MS}} &  \multicolumn{2}{c}{\textit{Spatial Unmatched JS}}\\
	\midrule
Lagos & 0.89 & (0.84, 0.94) & 0.88 & (0.83, 0.93) & 0.89 & (0.83, 0.93)\\
Ekiti & 0.87 & (0.79, 0.94) & 0.83 & (0.76, 0.89) & 0.83 & (0.75, 0.89)\\
Anambra & 0.80 & (0.73, 0.88) & 0.79 & (0.73, 0.85) & 0.79 & (0.72, 0.85)\\
Enugu & 0.80 & (0.71, 0.88) & 0.77 & (0.7, 0.85) & 0.77 & (0.69, 0.85)\\
Edo & 0.79 & (0.7, 0.88) & 0.78 & (0.7, 0.85) & 0.78 & (0.69, 0.85)\\
Osun & 0.77 & (0.69, 0.85) & 0.76 & (0.69, 0.83) & 0.76 & (0.68, 0.83)\\
Abia & 0.75 & (0.69, 0.82) & 0.75 & (0.69, 0.81) & 0.75 & (0.68, 0.82)\\
Delta & 0.75 & (0.69, 0.8) & 0.75 & (0.7, 0.8) & 0.76 & (0.69, 0.82)\\
Abuja & 0.73 & (0.68, 0.79) & 0.72 & (0.67, 0.77) & 0.72 & (0.65, 0.78)\\
Imo & 0.73 & (0.63, 0.84) & 0.74 & (0.65, 0.83) & 0.73 & (0.64, 0.82)\\
Bayelsa & 0.73 & (0.65, 0.8) & 0.73 & (0.66, 0.81) & 0.73 & (0.65, 0.81)\\
Ondo & 0.69 & (0.58, 0.8) & 0.70 & (0.6, 0.79) & 0.69 & (0.6, 0.78)\\
Rivers & 0.68 & (0.59, 0.77) & 0.69 & (0.61, 0.77) & 0.69 & (0.61, 0.77)\\
Cross River & 0.65 & (0.52, 0.78) & 0.66 & (0.55, 0.77) & 0.66 & (0.55, 0.76)\\
Adamawa & 0.65 & (0.57, 0.73) & 0.63 & (0.56, 0.71) & 0.63 & (0.55, 0.71)\\
Nassarawa & 0.65 & (0.54, 0.75) & 0.63 & (0.54, 0.72) & 0.63 & (0.54, 0.72)\\
Ebonyi & 0.63 & (0.57, 0.7) & 0.64 & (0.58, 0.7) & 0.64 & (0.58, 0.71)\\
Akwa Ibom & 0.63 & (0.55, 0.71) & 0.64 & (0.56, 0.72) & 0.64 & (0.56, 0.73)\\
Benue & 0.63 & (0.54, 0.71) & 0.63 & (0.55, 0.71) & 0.63 & (0.55, 0.7)\\
Oyo & 0.60 & (0.51, 0.7) & 0.61 & (0.52, 0.7) & 0.61 & (0.52, 0.7)\\
Plateau & 0.59 & (0.52, 0.65) & 0.58 & (0.52, 0.65) & 0.58 & (0.51, 0.65)\\
Kano & 0.56 & (0.5, 0.62) & 0.56 & (0.5, 0.62) & 0.56 & (0.5, 0.62)\\
Jigawa & 0.54 & (0.48, 0.6) & 0.53 & (0.48, 0.59) & 0.53 & (0.47, 0.59)\\
Kwara & 0.51 & (0.35, 0.67) & 0.55 & (0.42, 0.68) & 0.54 & (0.43, 0.66)\\
Ogun & 0.51 & (0.4, 0.62) & 0.55 & (0.45, 0.65) & 0.55 & (0.45, 0.66)\\
Borno & 0.49 & (0.42, 0.57) & 0.49 & (0.42, 0.56) & 0.49 & (0.41, 0.57)\\
Yobe & 0.45 & (0.4, 0.5) & 0.45 & (0.4, 0.5) & 0.45 & (0.39, 0.51)\\
Kaduna & 0.43 & (0.35, 0.5) & 0.43 & (0.36, 0.5) & 0.43 & (0.36, 0.5)\\
Kogi & 0.42 & (0.32, 0.53) & 0.49 & (0.39, 0.59) & 0.50 & (0.39, 0.6)\\
Taraba & 0.42 & (0.36, 0.48) & 0.43 & (0.37, 0.49) & 0.43 & (0.36, 0.5)\\
Niger & 0.39 & (0.27, 0.51) & 0.40 & (0.3, 0.51) & 0.41 & (0.32, 0.5)\\
Bauchi & 0.36 & (0.3, 0.43) & 0.37 & (0.31, 0.43) & 0.37 & (0.31, 0.44)\\
Katsina & 0.34 & (0.26, 0.41) & 0.34 & (0.27, 0.41) & 0.34 & (0.27, 0.41)\\
Kebbi & 0.31 & (0.25, 0.38) & 0.31 & (0.24, 0.37) & 0.31 & (0.24, 0.37)\\
Gombe & 0.28 & (0.2, 0.36) & 0.31 & (0.23, 0.38) & 0.31 & (0.24, 0.39)\\
Sokoto & 0.18 & (0.13, 0.23) & 0.18 & (0.13, 0.23) & 0.17 & (0.12, 0.23)\\
Zamfara & 0.14 & (0.07, 0.21) & 0.17 & (0.12, 0.23) & 0.18 & (0.12, 0.24)\\
\bottomrule
\end{tabular}

 		\caption{Point estimates of measles vaccination rates and 90\% interval estimates  for Admin-1 areas among children aged 12-23 months in Nigeria in 2018}
 		\label{tab:mcv-detail}
 	\end{table}
 	\begin{table}
 		\centering
 		\tiny
 		
\begin{tabular}{l|rl|rl|rl}
	\toprule
	State& \multicolumn{2}{c|}{\textit{H\'ajek}} &  \multicolumn{2}{c|}{\textit{Spatial Unmatched MS}} &  \multicolumn{2}{c}{\textit{Spatial Unmatched JS}}\\
	\midrule
Mulanje & 0.27 & (0.23, 0.32) & 0.25 & (0.21, 0.29) & 0.25 & (0.21, 0.29)\\
Blantyre & 0.24 & (0.2, 0.28) & 0.21 & (0.17, 0.26) & 0.22 & (0.18, 0.26)\\
Phalombe & 0.24 & (0.19, 0.29) & 0.23 & (0.19, 0.27) & 0.23 & (0.19, 0.28)\\
Zomba & 0.20 & (0.16, 0.24) & 0.19 & (0.16, 0.22) & 0.19 & (0.16, 0.22)\\
Ntcheu & 0.16 & (0.12, 0.21) & 0.15 & (0.11, 0.19) & 0.16 & (0.13, 0.2)\\
Mangochi & 0.16 & (0.12, 0.19) & 0.15 & (0.12, 0.18) & 0.15 & (0.12, 0.18)\\
Nsanje & 0.16 & (0.11, 0.2) & 0.15 & (0.11, 0.19) & 0.16 & (0.12, 0.2)\\
Thyolo & 0.15 & (0.11, 0.2) & 0.16 & (0.13, 0.19) & 0.16 & (0.13, 0.2)\\
Balaka & 0.15 & (0.11, 0.19) & 0.15 & (0.12, 0.18) & 0.15 & (0.12, 0.19)\\
Chiradzulu & 0.15 & (0.1, 0.19) & 0.16 & (0.12, 0.19) & 0.16 & (0.12, 0.2)\\
Neno & 0.14 & (0.1, 0.18) & 0.15 & (0.11, 0.18) & 0.15 & (0.12, 0.18)\\
Mwanza & 0.13 & (0.09, 0.17) & 0.14 & (0.1, 0.17) & 0.14 & (0.1, 0.17)\\
Karonga & 0.12 & (0.09, 0.15) & 0.11 & (0.08, 0.13) & 0.10 & (0.07, 0.13)\\
Chikwawa & 0.11 & (0.07, 0.15) & 0.13 & (0.1, 0.16) & 0.13 & (0.1, 0.16)\\
Nkhata Bay & 0.10 & (0.07, 0.14) & 0.09 & (0.06, 0.12) & 0.09 & (0.07, 0.12)\\
Nkhotakota & 0.10 & (0.07, 0.13) & 0.09 & (0.07, 0.11) & 0.09 & (0.07, 0.11)\\
Machinga & 0.09 & (0.07, 0.12) & 0.11 & (0.08, 0.13) & 0.10 & (0.08, 0.13)\\
Kasungu & 0.09 & (0.06, 0.12) & 0.08 & (0.06, 0.11) & 0.09 & (0.07, 0.11)\\
Rumphi & 0.09 & (0.05, 0.12) & 0.08 & (0.06, 0.11) & 0.09 & (0.06, 0.11)\\
Lilongwe & 0.09 & (0.06, 0.11) & 0.08 & (0.06, 0.11) & 0.08 & (0.06, 0.11)\\
Dedza & 0.08 & (0.04, 0.12) & 0.09 & (0.07, 0.12) & 0.10 & (0.08, 0.13)\\
Mchinji & 0.08 & (0.06, 0.1) & 0.08 & (0.06, 0.1) & 0.07 & (0.05, 0.09)\\
Salima & 0.07 & (0.05, 0.1) & 0.08 & (0.06, 0.1) & 0.07 & (0.06, 0.1)\\
Ntchisi & 0.07 & (0.05, 0.1) & 0.07 & (0.06, 0.09) & 0.07 & (0.05, 0.09)\\
Dowa & 0.07 & (0.04, 0.1) & 0.07 & (0.05, 0.09) & 0.07 & (0.05, 0.09)\\
Chitipa & 0.06 & (0.03, 0.09) & 0.07 & (0.05, 0.09) & 0.06 & (0.04, 0.09)\\
Mzimba & 0.06 & (0.04, 0.08) & 0.06 & (0.05, 0.08) & 0.06 & (0.05, 0.08)\\
\bottomrule
\end{tabular}

 		\caption{Point estimates of HIV prevalence and 90\% interval estimates  for Admin-1 areas among women aged 15-29 in Nigeria in 2018}
 		\label{tab:hiv-detail}
 	\end{table}
\end{document}